# Room-temperature and tunable tunneling magnetoresistance in $Fe_3GaTe_2$-based all-2D van der Waals heterojunctions with high spin polarization

*Wen Jin[1,2], Gaojie Zhang[1,2], Hao Wu[1,2,3], Li Yang[1,2], Wenfeng Zhang[1,2,3], Haixin Chang[1,2,3],\**

[1]Center for Joining and Electronic Packaging, State Key Laboratory of Material Processing and Die & Mold Technology, School of Materials Science and Engineering, Huazhong University of Science and Technology, Wuhan 430074, China.

[2]Shenzhen R&D Center of Huazhong University of Science and Technology (HUST), Shenzhen 518000, China.

[3]Wuhan National High Magnetic Field Center and Institute for Quantum Science and Engineering, Huazhong University of Science and Technology, Wuhan 430074, China.

\*Corresponding author. E-mail: hxchang@hust.edu.cn

## Abstract

Magnetic tunnel junctions (MTJs) based on all-two dimensional (2D) van der Waals heterostructures with sharp and clean interfaces in atomic scale are essential for the application of next-generation spintronics. However, the lack of room-temperature intrinsic ferromagnetic crystals with perpendicular magnetic anisotropy has greatly hindered the development of vertical MTJs. The discovery of room-temperature intrinsic ferromagnetic 2D crystal $Fe_3GaTe_2$ has solved the problem and greatly facilitated the realization of practical spintronic devices. Here, we demonstrate a room-temperature MTJ based on $Fe_3GaTe_2/WS_2/Fe_3GaTe_2$ heterostructure. The tunnelling magnetoresistance (TMR) ratio is up to 213% with high spin polarization of 72% at 10 K, the highest ever reported in $Fe_3GaTe_2$-based MTJs up to now. The tunnelling spin-valve signal robustly exists at room temperature (300 K) with bias current down to 10 nA. Moreover, the spin polarization can be modulated by bias current and the TMR shows a sign reversal at large bias current. Our work sheds light on the potential application for low-energy consumption all-2D vdW spintronics and offers alternative routes for the electronic control of spintronic devices.

**Main text**

Magnetic tunnel junctions (MTJs) are crucial building blocks for spintronics owing to their nonvolatile high- and low-resistance states controlled by different magnetic fields and have been intensively explored as magnetic sensors, magnetic random-access memories and magnetic logic gates [1-5]. In principle, MTJs are consisted of two ferromagnetic electrodes and a non-magnetic interlayer. Various non-magnetic materials have been applied to realize such heterojunction acting as tunnel barriers, such as $Al_2O_3$ [6], MgO [7], BN [8] and, recently, transition-metal dichalcogenides (TMDCs) [9-11]. As two-dimensional (2D) van der Waals (vdW) materials, the weak interlayer force and atomically sharp surfaces without dangling bonds [12] of TMDCs provide platforms to integrate heterostructure with ferromagnets and realize high spin polarization of the magnetic materials.

Conventional ferromagnets, such as $Fe_3O_4$ [9], NiFe [13], Co [14], have been applied to integrate with TMDCs to fabricate spin-valve devices including MTJs. But the MR ratio only ranges from 0.2%-3.2% at low temperature (10-80 K), which can be attributed to the degradation of the interface caused by the inevitable damage of the TMDCs during metal deposition. Thus, it is highly desirable to assemble 2D ferromagnets with TMDCs to realize seamless heterostructure with well-defined interface. Through the efforts of recent years, experimental results have proved that ferromagnetism can be maintained in 2D system by magnetic anisotropy, which opens a spin-wave excitation gap that effectively suppress thermal fluctuations [15-18]. 2D vdW ferromagnetic crystals with perpendicular magnetic anisotropy (PMA) facilitates the fabrication of all-2D vdW spin valves and MTJs with high-quality interfaces due to their easy-exfoliation nature and atomically flat surface. For example, MTJs based on $Fe_3GeTe_2$ (FGT) /$WSe_2$/FGT [19], FGT/InSe/FGT [20] and FGT/hBN/FGT [21] heterojunctions demonstrated TMRs of 25.8%-300% at low temperature (10 K or 4.2 K), but none of the TMRs were preserved at room temperature due to the low Curie temperature of FGT (~220 K) [17], which, greatly hindered the practical application in next-generation spintronic devices. The recent discovery of above-room-temperature intrinsic ferromagnetic 2D vdW crystal $Fe_3GaTe_2$ with large PMA and high saturation magnetic moment has shed light on the fabrication of room-temperature MTJs for practical use [22]. Despite that room-temperature spin valves including $Fe_3GaTe_2$/$MoS_2$/$Fe_3GaTe_2$ [23,24], $Fe_3GaTe_2$/$WSe_2$/$Fe_3GaTe_2$ [25], and $Fe_3GaTe_2$/$MoSe_2$/$Fe_3GaTe_2$ [26] have been reported, the research on room-temperature $Fe_3GaTe_2$-based MTJs with $WS_2$ barrier is still rare.

Here, we realize room-temperature MTJs based on vertical all-2D vdW heterostructures using $Fe_3GaTe_2$ as the top and down ferromagnetic electrodes and $WS_2$ as the barrier. The non-linear I-V curves manifest the tunnelling transport behavior of the $Fe_3GaTe_2/WS_2/Fe_3GaTe_2$ heterojunction. A TMR ratio up to 213% and an estimated high spin polarization of 72% are obtained in the MTJ at low temperature (10 K), the highest amongst all-2D vdW $Fe_3GaTe_2$-based MTJs. With increasing temperature, the TMR ratio decreases but still retains its value up to 11% at room temperature. Moreover, the polarity of the TMR can be artificially modulated by the bias current, varying from +213% to -9% with increasing bias, demonstrating the spin-filtering effect.

Figure 1(a) depicts the schematic diagram of the $Fe_3GaTe_2/WS_2/Fe_3GaTe_2$ MTJ structure, where 2D $WS_2$ layer is sandwiched by the top and bottom $Fe_3GaTe_2$ ferromagnetic layers, forming a vertical heterostructure. To protect the heterojunction from inevitable damage and contamination during the metal deposition, the whole junction was encapsuled by h-BN (fabrication details in Supplementary Note 1). The optical image of the heterojunction is shown in the upper-left inset of Fig.1(b). Atomic force microscopy (AFM) characterization was performed to determine the thickness of the device, indicating the bottom $Fe_3GaTe_2$, $WS_2$ interlayer and top $Fe_3GaTe_2$ are 11, 4.2, and 12 nm, respectively [Fig.1(b)]. Figure S1(a) shows the front view of the schematic crystal structure of $Fe_3GaTe_2$, where $Fe_3Ga$ slabs are sandwiched between the top and bottom Te layers. In 2D $Fe_3GaTe_2$, the adjacent layers are stacked by vdW coupling with the interlayer spacing of 0.78 nm. To identify the magneto-transport properties of the 2D $Fe_3GaTe_2$ single crystal, a 13 nm-thick $Fe_3GaTe_2$ Hall bar device was fabricated and measured with magnetic field applied perpendicular to the device. Figure S1(b) shows the longitudinal resistance $R_{xx}$ as a function of temperature, where $R_{xx}$ decreases monotonously as temperature declines, indicating the metallic behavior of $Fe_3GaTe_2$. The anomalous Hall resistance ($R_{xy}$) versus magnetic field curves measured at various temperatures ranging from 10 to 350 K are shown in Fig. S1(c). With increasing temperature, the coercivity of the $Fe_3GaTe_2$ declines as a consequence of thermal fluctuation enhancement and finally vanishes, showing a $T_C$ over ~320 K, which is in good agreement with our previous report. Meanwhile, the nearly rectangular hysteresis loop implies a perpendicular magnetic anisotropy, making $Fe_3GaTe_2$ suitable for utilizing in vertical MTJs. Figure S2 presents the Raman spectra of the interlayer $WS_2$ (4.2 nm), where the two main peaks of $E^1_{2g}$ and $A_{1g}$ are located at ~350 $cm^{-1}$ and ~420 $cm^{-1}$, respectively, similar to the previous report [27].

We first investigated the electrical characteristics of the $Fe_3GaTe_2/WS_2/Fe_3GaTe_2$ heterojunction. As shown in Fig.2(a), the symmetric and nonlinear current versus voltage (I-V) curves at 10 K and 300 K exhibit a typical tunneling behavior. Then we conducted the magneto-transport measurements with an out-of-plane magnetic field to control the magnetization direction of the top and bottom $Fe_3GaTe_2$. The magnetoresistance versus magnetic field (R-B) curve with 10 nA bias current at 10 K is shown in Fig.2(b), where a typical large tunnel spin-valve effect was observed. As the external magnetic field ramping from -1.5 T to 1.5 T for a forward sweep (purple square), a sharp increase of the magnetoresistance is observed at B~0.56 T and maintained until B~0.8 T, which corresponds to the high-resistance $R_{AP}$, implying the magnetization directions of the two $Fe_3GaTe_2$ ferromagnetic electrodes are in an antiparallel state. Then the magnetoresistance jumps back to the relatively small value as B>0.8 T, which corresponds to the low-resistance $R_P$, implying the magnetization directions of the two $Fe_3GaTe_2$ ferromagnetic electrodes are in a parallel state. When the external magnetic field sweeps back from 1.5 T to -1.5 T (green square), the resistance shows similar up and down switching. The bistable $R_{AP}$ and $R_P$ indicate that the overall tunnel spin-valve operations are controlled by the magnetization switching of the top and bottom $Fe_3GaTe_2$. The TMR ratio can be calculated as $TMR = (R_{AP} - R_P)/R_P$. According to the equation, the TMR ratio of the device is estimated to be 173%, while another device with thinner $WS_2$ spacer (3.5 nm) exhibits a larger TMR of 213% at the same temperature (10 K) [Fig. S4 (b)], showing the thickness-dependence behavior. The TMR ratio can be further increased by applying thinner tunnel barrier.

To characterize the temperature dependence of the device, we further performed the magneto-transport measurements at various temperatures from 10 to 300 K. In Figs. 3(a) and (b), the TMR ratios exhibit a downward transition with increasing temperature and still retains 11% at room temperature. The increasing temperature intensifies the thermal fluctuation, resulting in the reduction of perpendicular anisotropic energy and coercivity of $Fe_3GaTe_2$, thus, contributes to the smaller TMR. Since both of the top and bottom FM electrodes are adopting the same ferromagnetic crystal Fe$_3$GaTe$_2$, the relationship between the TMR ratio and spin polarization of Fe$_3$GaTe$_2$ can be described as $TMR = \frac{2P^2}{1-P^2}$, based on the modified Julliere's model [28]. Thus, the spin polarization $P$ of the Fe$_3$GaTe$_2$ ferromagnetic electrodes can be estimated as 68% (TMR=173%, 10 K). A higher $P$ is obtained from another device at the value of 72% (TMR=213%, 10 K) [Fig. S5 (c)]. The $P$ of Fe$_3$GaTe$_2$ is much larger than that of most reported 2D materials sandwiched by conventional 3D FM electrodes such as Co/graphene/Co (TMR=1%, $P$=7.1%, 3 K) [29] and NiFe/MoS$_2$/NiFe [13] (TMR=0.73%, $P$=6.2%, 20 K) (Fig.S6). Compared to other all-2D vdWs heterostructures (Fig.S6), this value is also higher than that of FGT/h-BN/FGT (TMR=160%, P=66%, 4.2K) [31], FGT/InSe/FGT (TMR=41%, $P$=41%, 10 K) [20], Fe$_3$GaTe$_2$/MoS$_2$/Fe$_3$GaTe$_2$ (TMR=15.89%, $P$=27% 10 K) [23,24], Fe$_3$GaTe$_2$/MoSe$_2$/Fe$_3$GaTe$_2$ (TMR=37.7%, $P$=39.8%, 2 K) [26], and is comparable to that of Fe$_3$GaTe$_2$/WSe$_2$/Fe$_3$GaTe$_2$ (TMR=210%, $P$=71.5%, 10 K) [25]. However, this value is still lower than that of FGT/h-BN/ FGT (TMR=300%, P=77%, 4.2 K) [21] due to the wider band-gap and better spin-preservation abilities of h-BN. The temperature evolution of the $P$ magnitude can be described by the equation of $P(T) = P(0)(1 - \alpha T^{\frac{3}{2}})$ based on the Bloch's law, where $P(0)$ and α refer to the spin polarization at 0 K and material-dependent constant, respectively [35,36]. The α of our device is calculated to be 1.17×10$^{-4}$K$^{-3/2}$, which is comparable to the value reported in other literature [20,23-26].

The measurement of TMR ratio with bias current ranging from 10 nA to 15 μA was conducted at 10 K. As shown in Fig.4 (a), stable tunnel spin-valve signals are observed at low bias current down to 10 nA, suggesting that the MTJ can work as a low-energy consumption device. Figure 4 (c) is the TMR versus bias current point plot extracted from Fig.4 (a). When reversing the direction of the bias current, the magnitude of the TMR ratios show little difference, which is attributed to the symmetrical interface of the $Fe_3GaTe_2/WS_2/Fe_3GaTe_2$ junction. We note that the I-symmetric TMR ratio decreases rapidly with increasing |I| at first and drops approximately to zero at I ~3 μA. As the bias current increases further, the TMR ratio turns to be negative and reaches to -3% at I=4 μA. With the continuously rising bias current, the polarity of the TMR keeps negative and the ratio reaches its maximum and maintains at ~-4%. This intriguing phenomenon is also observed in another device with TMR ratio varying in a wide range from +213% to -9%, while the TMR ratio reverses its polarity at I>10 μA and reaches its maximum of -9% at I=20 μA (Fig. S7). Similarly, as the bias current keeps rising, the TMR ratio keeps negative and maintains at~9%. Figure 5 shows the schematic diagrams of the electrons transport in $Fe_3GaTe_2/WS_2/Fe_3GaTe_2$ MTJs to explain the mechanism of the TMR polarity reversal. In the MTJ, the tunnel spin-valve effect is based on the difference in the density of states (DOS) around the Fermi level ($E_F$) of the ferromagnetic electrodes between spin-up and spin-down electrons [37]. When the two ferromagnetic electrodes are in the parallel configuration, the electrons in the majority-spin sub-band and minority-spin sub-band of ferromagnetic metal 1 (FM1) will tunnel into the empty states in the majority-spin sub-band and minority-spin sub-band of ferromagnetic metal 2 (FM2) during the magneto-transport process, respectively, yielding a low junction resistance ($R_P$) [Fig.5 (a) i]. When the two FMs are in the antiparallel configuration, the majority electrons of FM1 will tunnel into the empty states in the minority-spin sub-band of FM2, while the minority electrons of FM1 will tunnel into the empty states in the majority-spin sub-band of FM2, as a consequence, the quantity of the transport electrons decreases, leading to a high junction resistance ($R_{AP}$) [Fig.5 (a) ii]. Normally, $R_{AP}$ is larger than $R_P$, contributing to a positive TMR ratio. However, the spin states can be modulated by the applied large bias current [Fig.5 (b)]. Similar to FGT, $Fe_3GaTe_2$ owns an electronic structure consisted of itinerant and localized spin states [38], which respectively determine the spin current and magnetization of $Fe_3GaTe_2$. The spin polarization can be tuned by the localized spin states with high energy [21], in particular, by applying a large bias current, an energy window of the electronic states will be opened, the $E_F$ of FM1 will align to the empty states of the biased FM2, and the $E_F$ of the biased FM2 will align to the filled states of FM1, thus, reverse the TMR ratio to negative ($R_{AP} < R_P$). Our work offers an opportunity for more delicate modulation of the spin-electrons polarization through electronic control in all-2D vdW MTJs, making it more appealing to apply MTJs into next-generation spintronics applications.

In summary, we report a room-temperature $Fe_3GaTe_2$/$WS_2$/$Fe_3GaTe_2$ MTJ. A TMR ratio up to 213% is observed at 10 K and up to 11% at room temperature with modulating bias current down to 10 nA. The deduced spin polarization is 72% at 10 K, the highest amongst $Fe_3GaTe_2$-based all-2D vdW spin valves. The TMR ratio shows strong dependence on thickness of the $WS_2$ tunnel barrier. Moreover, the TMR polarity can be reversed by modulating the applied bias current and a negative TMR emerges at large bias current, which, is contributed from the localized spin states with high-energy in $Fe_3GaTe_2$. Our work indicates the potential application for next-generation all-2D vdW spintronics with low-energy consumption and opens up alternative routes for the electronic control of the spin in 2D devices.

## Acknowledgments

This work was supported by the National Key Research and Development Program of China (No. 2022YFE0134600) and the National Natural Science Foundation of China (No. 52272152, 61674063 and 62074061), the Foundation of Shenzhen Science and Technology Innovation Committee (JCYJ20210324142010030 and JCYJ20180504170444967), and the fellowship of China Postdoctoral Science Foundation (No. 2022M711234). We thank the AFM and Raman tests from Analytical Center of Huazhong University of Science and Technology.

## AUTHOR DECLARATIONS

### Conflict of Interest

The authors have no conflicts to disclose.

## Author contributions

Wen Jin: Investigation (lead); Methodology (lead). Gaojie Zhang: Methodology (supporting). Hao Wu: Investigation (supporting). Li Yang: Methodology (supporting). Wenfeng Zhang: Project administration (supporting); Supervision (supporting). Haixin Chang: Conceptualization (lead); Funding acquisition (lead); Project administration (lead); Validation (lead); Writing-review and editing (lead).

## DATA AVAILABILITY

The data that support the findings of this study are available within the article and its supplementary material.

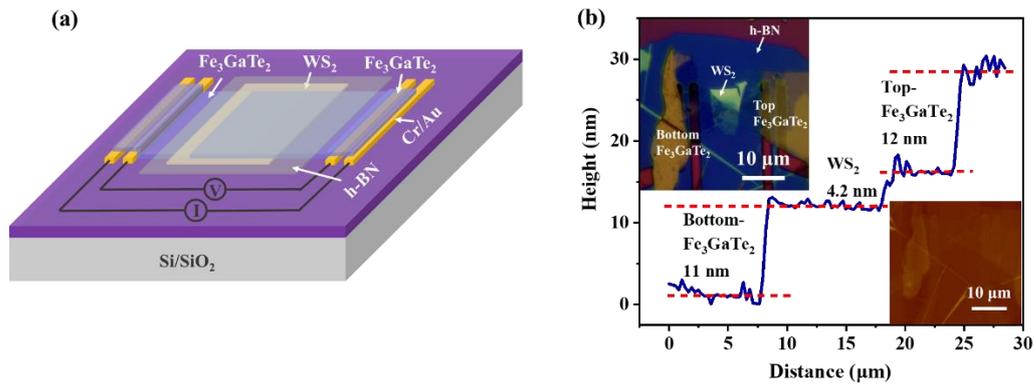

**FIG.1.** (a) Schematic diagram of the $Fe_3GaTe_2/WS_2/Fe_3GaTe_2$ MTJs. (b) AFM characterization of a typical device. The thickness of the bottom $Fe_3GaTe_2$, $WS_2$ and top $Fe_3GaTe_2$ are 11, 4.2 and 12 nm, respectively. Inset: the optical (upper left) and AFM images (bottom right) of the MTJ.

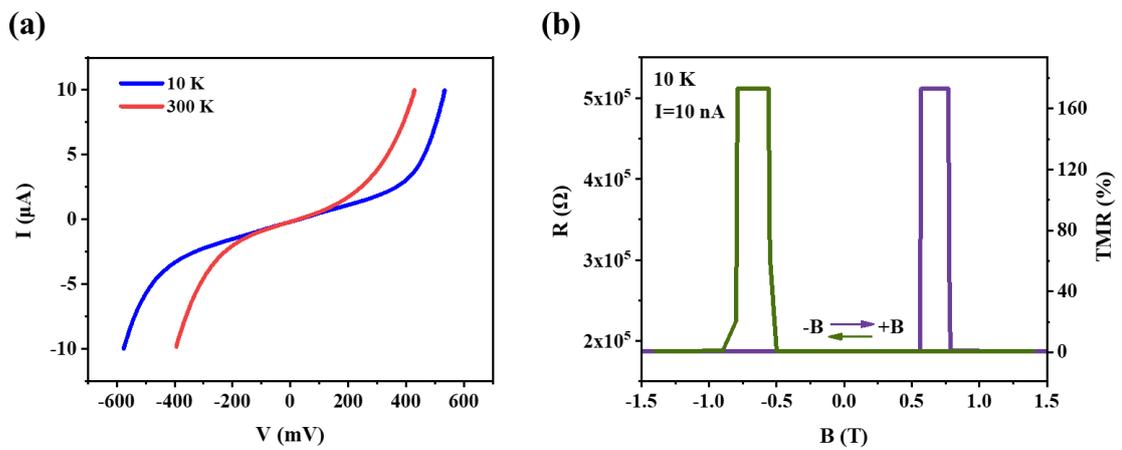

**FIG.2.** (a) I-V characteristics at 10 K and 300 K. (b) Resistance and TMR vs perpendicular magnetic field at 10 K with a bias current of 10 nA.

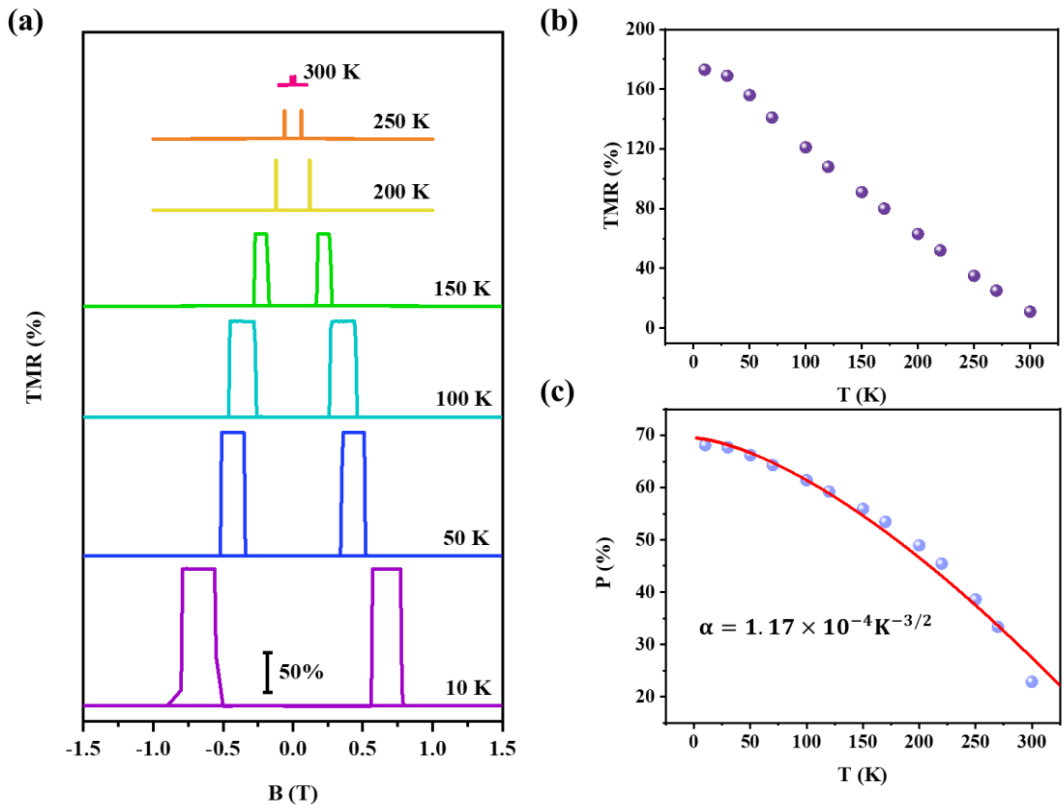

**FIG.3.** (a) TMR curves measured at various temperatures ranging from 10 K to 300 K with a fixed bias current of 10 nA. (b) The temperature-dependent TMR ratios extracted from Fig.3(a). (c) The temperature-dependent spin polarizations deduced from TMR ratios. The red line is the fitting curve according to Bloch's law.

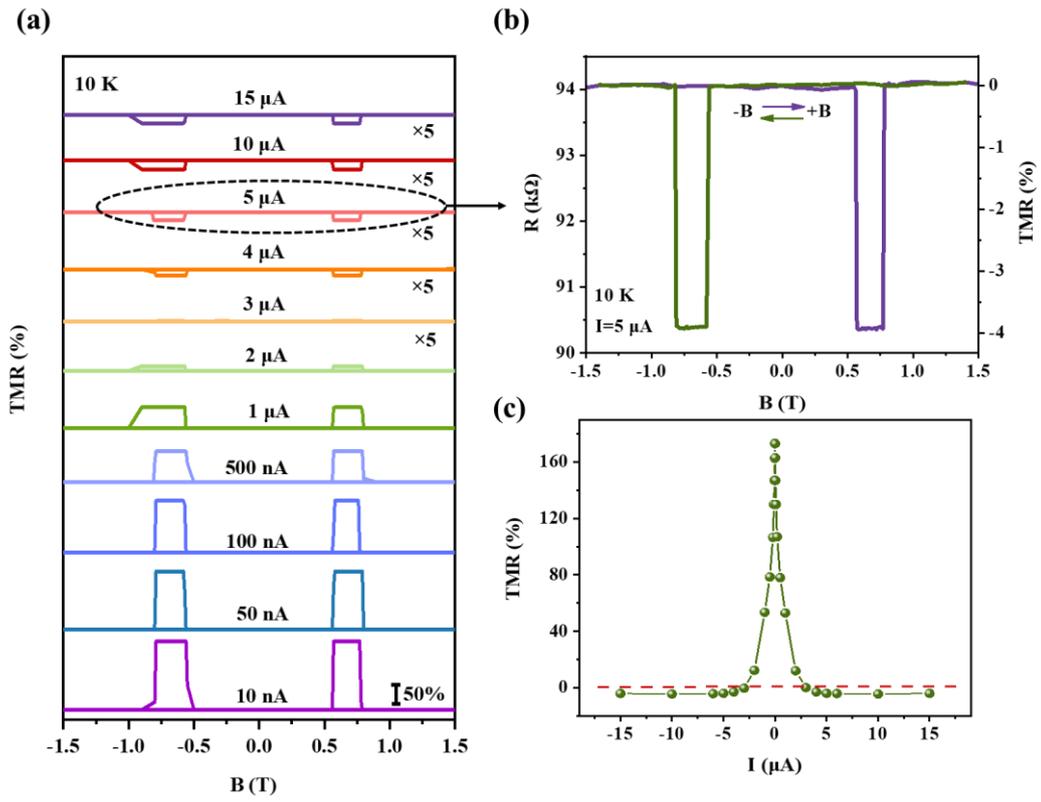

**FIG.4.** (a) TMR curves measured at various bias current ranging from 10 nA to 15 μA at 10 K. (b) The magnified TMR curve from Fig.4(a) measured with bias current of 5 μA. (c) The I-dependent TMR ratio variations for the $Fe_3GaTe_2/WS_2/Fe_3GaTe_2$ MTJ.

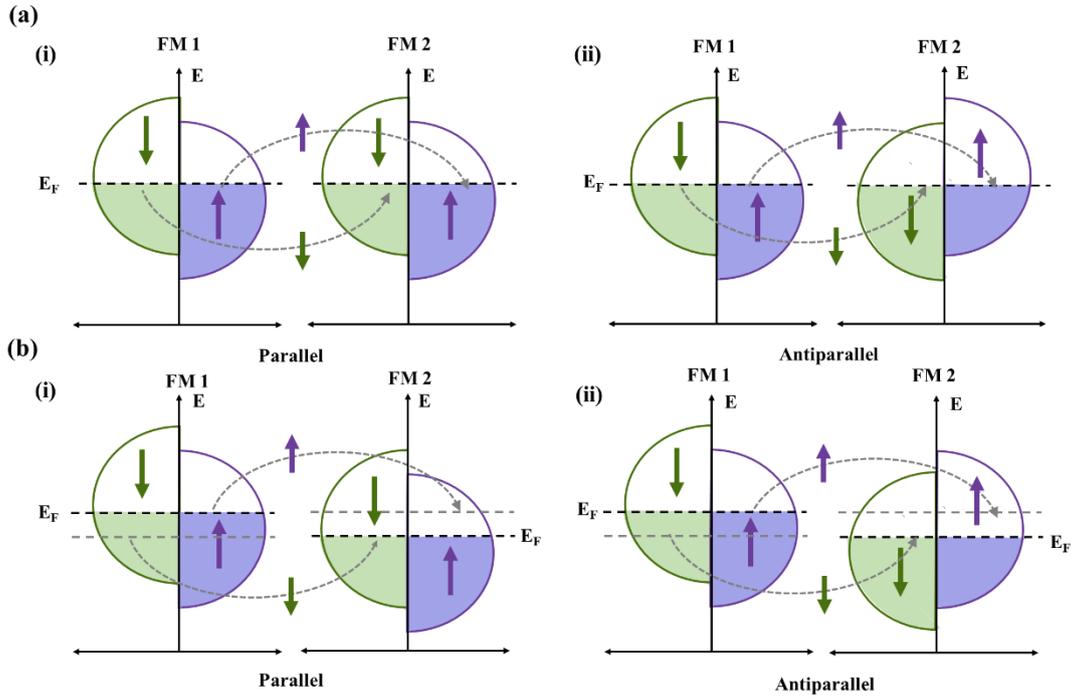

**FIG.5.** (a) Schematic diagrams of the electrons transport in Fe$_3$GaTe$_2$/WS$_2$/Fe$_3$GaTe$_2$ MTJs, when the two Fe$_3$GaTe$_2$ electrodes are in the parallel (i) and antiparallel (ii) configuration, respectively. (b) Schematic diagrams of the electrons transport in Fe$_3$GaTe$_2$/WS$_2$/Fe$_3$GaTe$_2$ MTJs with large bias current, when the two Fe$_3$GaTe$_2$ electrodes are in the parallel (i) and antiparallel (ii) configuration, respectively.

Supporting Information

# Room-temperature and tunable tunneling magnetoresistance in Fe$_3$GaTe$_2$-based all-2D van der Waals heterojunctions with high spin polarization

*Wen Jin[1,2], Gaojie Zhang[1,2], Hao Wu[1,2,3], Li Yang[1,2], Wenfeng Zhang[1,2,3], Haixin Chang[1,2,3, *]*

[1]Center for Joining and Electronic Packaging, State Key Laboratory of Material Processing and Die & Mold Technology, School of Materials Science and Engineering, Huazhong University of Science and Technology, Wuhan 430074, China.

[2]Shenzhen R&D Center of Huazhong University of Science and Technology (HUST), Shenzhen 518000, China.

[3] Wuhan National High Magnetic Field Center and Institute for Quantum Science and Engineering, Huazhong University of Science and Technology, Wuhan 430074, China.

[*]Corresponding author. E-mail: hxchang@hust.edu.cn

## Supplementary Note 1: Methods

**Device fabrication**

**$Fe_3GaTe_2/WS_2/Fe_3GaTe_2$ MTJs:** The 2D vdW crystal $Fe_3GaTe_2$ was grown by a self-flux method, $WS_2$ and hBN were purchased from HQ Graphene. To begin with, four-terminal electrodes were pre-patterned on a $SiO_2/Si$ substrate by laser direct writing machine (Micro Writer ML3, DMO). Followed by a metal deposition process with Cr/Au of 5/25 nm and a lift-off procedure. With the help of optical microscope, $Fe_3GaTe_2$ flakes with appropriated thickness and shape was chosen and transferred onto the pre-fabricated electrodes on the $SiO_2/Si$ substrate by a site-controllable dry transfer method. Then, a $WS_2$ flake were transferred onto the $Fe_3GaTe_2$ flake using the same method, followed by another transfer of $Fe_3GaTe_2$ flake with different thickness and shape to fabricate an all-2D vdW heterojunction. A BN flake was transferred at last to cover the whole junction area to prevent the device from oxidation. The whole procedure was performed in the nitrogen-filled glove box ($H_2O$, $O_2$<0.1 ppm).

**$Fe_3GaTe_2$ Hall device:** Six-terminal Hall bar electrodes with Cr/Au of 5/25 nm was pre-deposited on a $SiO_2/Si$ substrate using the similar process. The $Fe_3GaTe_2$ flake was mechanically exfoliated and transferred onto the Hall bar by using the polydimethylsiloxane (PDMS) stamps. The whole procedure was performed in the nitrogen-filled glove box ($H_2O$, $O_2$<0.1 ppm).

**Device characterization**

The thickness of the $Fe_3GaTe_2$ and $WS_2$ flakes were characterized by atomic force microscopy (AFM, XE7, Park; SPM9700, Shimadzu; Dimension EDGE, Bruker). The electrical transport and magneto-transport properties were measured in a physical property measurement system (PPMS, DynaCool, Quantum Design). The magnetic field was applied perpendicular to the device.

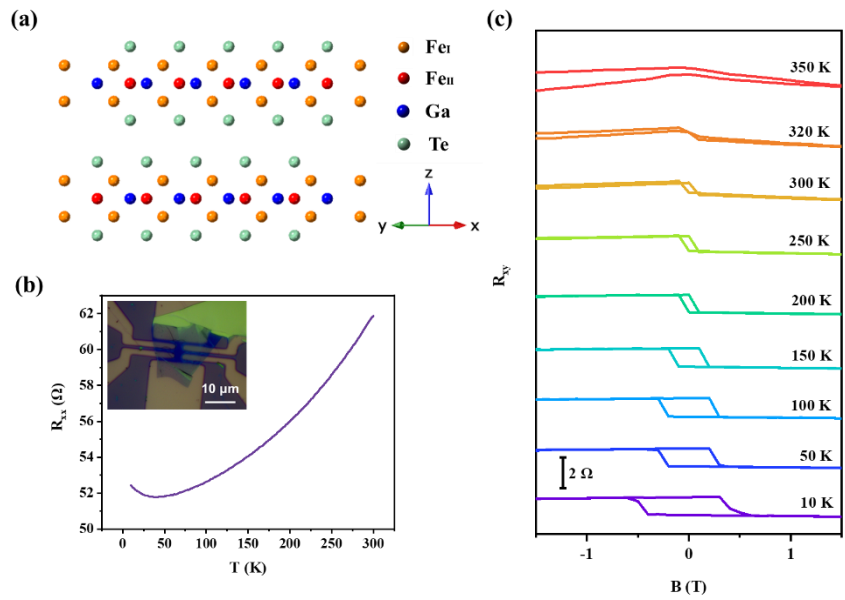

**FIG.S1.** (a) Front view of the crystal structure of $Fe_3GaTe_2$. (b) Temperature dependence of the longitudinal resistance ($R_{xx}$). Inset: optical image of the 13 nm-thick $Fe_3GaTe_2$ Hall device. (c) Hall resistance ($R_{xy}$) at different temperatures from 10 K to 350 K.

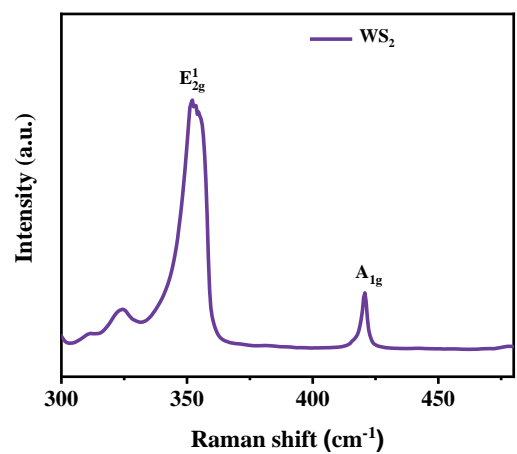

**FIG.S2.** Raman spectra of $WS_2$ (4.2 nm).

# Supplementary Note 2: characterization, electrical and magneto-transport properties of another device with thinner WS₂ barrier

Figs. S3-S5 and S6 show the characterization, electrical properties and magneto-transport properties of another device. The AFM characterization shows that the $WS_2$ spacer layer of the $Fe_3GaTe_2/WS_2/Fe_3GaTe_2$ heterojunction is ~3.5 nm, thinner than that in the main text (~4.2 nm). As a consequence, the MTJ exhibits a larger TMR of 213% at 10 K. However, the TMR of this device at room temperature (7%, 300 K) is smaller than that of the device in the main text (11%, 300 K), which is attributed to difference in $T_C$ of FGT between the two devices.

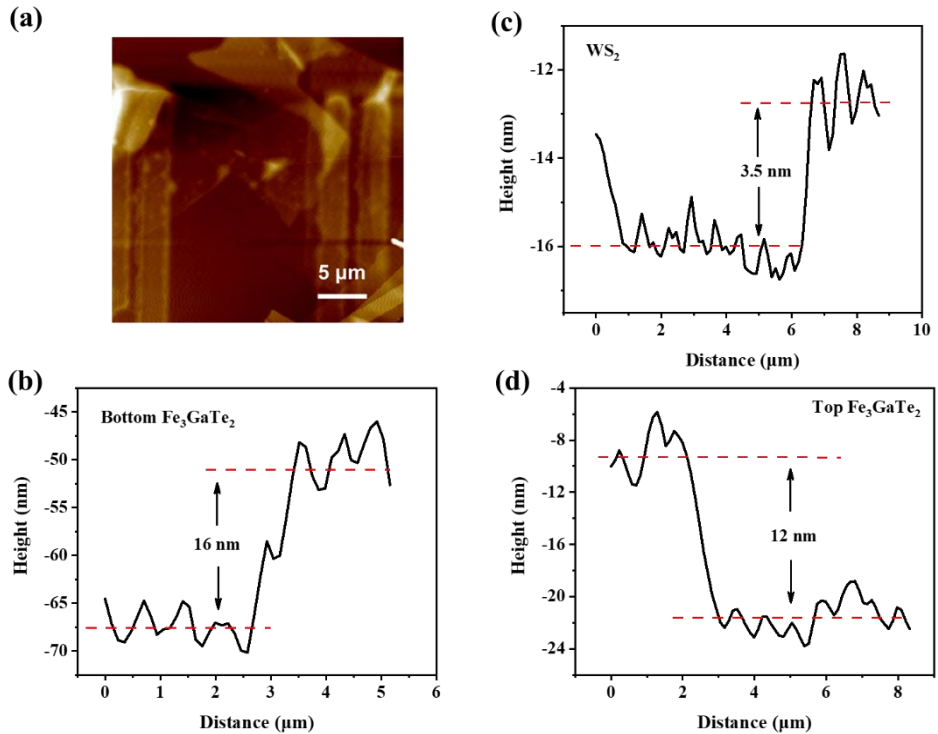

**FIG.S3.** (a) AFM image of another $Fe_3GaTe_2/WS_2/Fe_3GaTe_2$ MTJ device with a thiner (3.5 nm) spacer layer. (b)-(d) The height profile of the bottom $Fe_3GaTe_2$, $WS_2$ and top $Fe_3GaTe_2$ layer, indicating the thickness of 16, 3.5 and 12 nm, respectively.

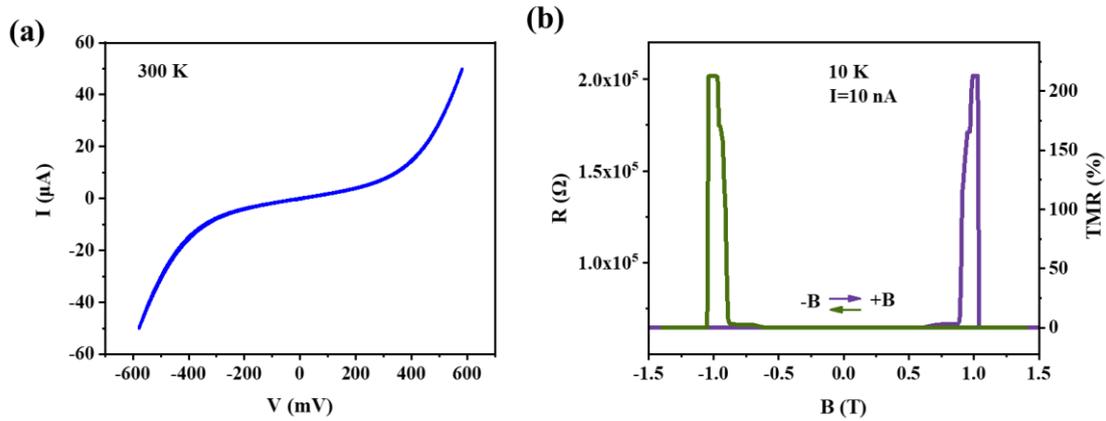

**FIG.S4.** (a) I-V characteristics at 300 K. (b) Resistance and TMR vs perpendicular magnetic field at 10 K with a bias current of 10 nA.

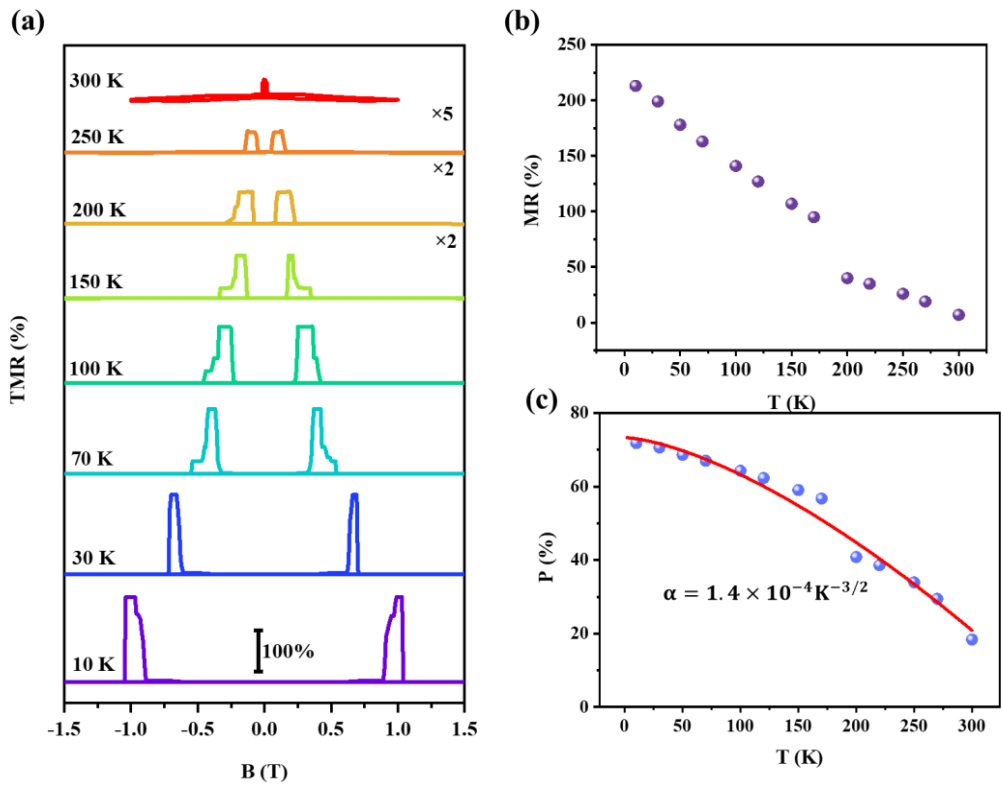

**FIG.S5.** (a) TMR curves for 3.5 nm WS$_2$ MTJ measured at various temperatures ranging from 10 K to 300 K with a fixed bias current of 10 nA. (b) The extracted temperature-dependent TMR ratios. (c) The temperature-dependent spin polarizations. The red line is the fitting curve according to Bloch's law.

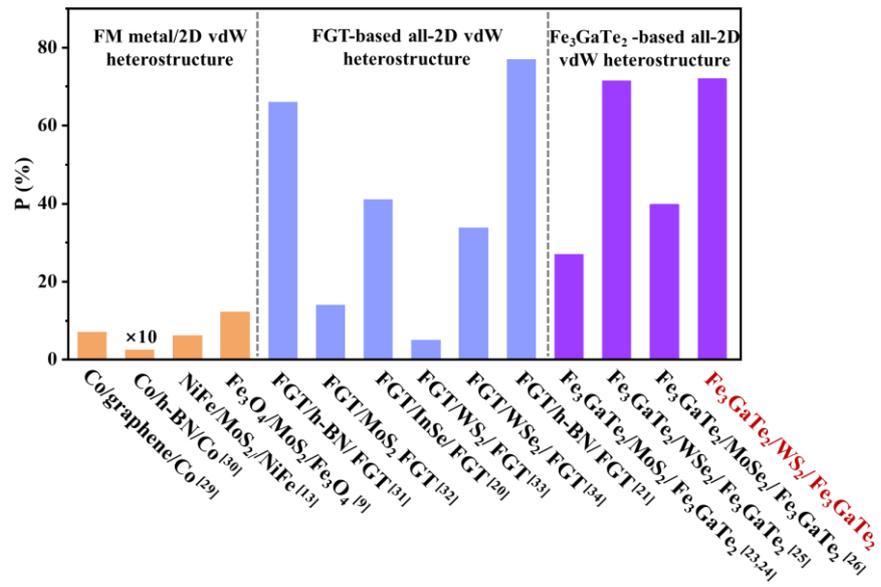

**FIG.S6.** Comparison of the spin polarization of FM metal/2D vdW heterostructure, FGT-based all-2D vdW heterostructure and $Fe_3GaTe_2$-based all-2D vdW heterostructure from previous reports [9,13,20,21,23-26,29-32].

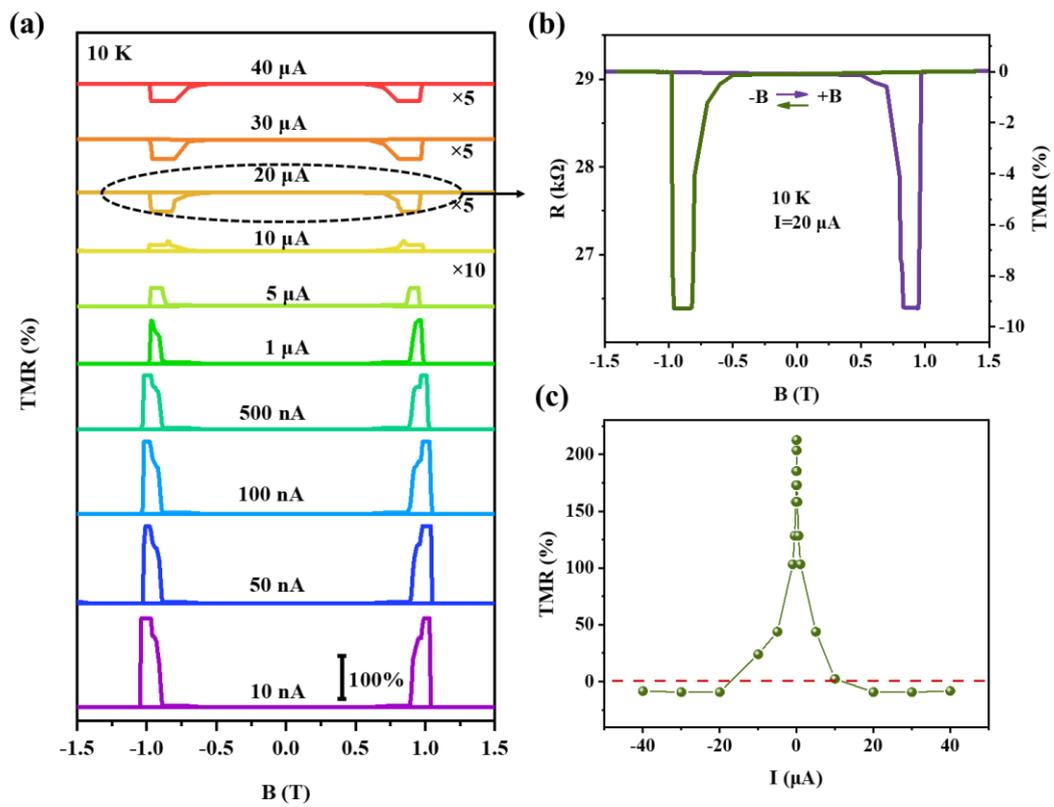

**FIG.S7.** (a) TMR curves measured at various bias current ranging from 10 nA to 40 µA at 10 K with 3.5 nm-thick $WS_2$ barrier. (b) The magnified TMR curve from Fig.S7 (a) measured with bias current of 20 µA. (c) The I-dependent TMR ratio variations for the $Fe_3GaTe_2/WS_2/Fe_3GaTe_2$ MTJ with 3.5 nm-thick $WS_2$ barrier.